\documentclass[pdflatex,sn-mathphys-num]{sn-jnl}

\usepackage{graphicx}%
\usepackage{multirow}%
\usepackage{amsmath,amssymb,amsfonts}%
\usepackage{amsthm}%
\usepackage{mathrsfs}%
\usepackage[title]{appendix}%
\usepackage{xcolor}%
\usepackage{textcomp}%
\usepackage{manyfoot}%
\usepackage{booktabs}%
\usepackage{algorithm}%
\usepackage{algorithmicx}%
\usepackage{algpseudocode}%
\usepackage{listings}%
\usepackage{bm}
\usepackage{mathrsfs}
\usepackage{ulem}

\theoremstyle{thmstyleone}%
\theoremstyle{thmstyletwo}%

\theoremstyle{thmstylethree}%

\begin{document}

\title[Article Title]{N-fold topological mode replication in hierarchical honeycomb lattices
}


\author*{\fnm{Keita} \sur{Funayama}}\email{funayama@mosk.tytlabs.co.jp}

\author{\fnm{Kenichi} \sur{Yatsugi}}

\author{\fnm{Hideo} \sur{Iizuka}}

\affil{\orgname{Toyota Central R\&D Labs, Inc.}, \orgaddress{\street{41-1 Yokomichi}, \city{Nagakute}, \postcode{480-1192}, \state{Aichi}, \country{Japan}}}


\abstract{
Multi-band topological states enable robust and versatile wave manipulation across a variety of physical platforms.
However, the emergence of multi-band topological states has relied on higher-frequency modes with complex spatial profiles, which constrains the realization of robust topological states due to fragile symmetry and pseudospin hybridization in these modes.
Here, we show a general design principle for scalable multi-band topological states by replicating a robust fundamental topological mode in the frequency domain.
By introducing hierarchical resonators as an internal degree of freedom into a quantum spin Hall-based lattice, multiple topological states emerge discretely in correspondence with the hierarchical levels while preserving the spatial profile of the fundamental mode at the host lattice.
Implementing this design principle in a versatile microelectromechanical platform, we experimentally demonstrate that the fundamental and replicated topological modes propagate simultaneously in a single waveguide while suppressing mutual cross-talk.
Our results establish topology replication as a universal strategy for designing multi-band topological systems and open routes toward multi-channel topological wave devices. 
}

\keywords{mode replication, hierarchical structure, topological phononic crystal, multi-channel waveguide}

\maketitle

\section{Introduction}\label{sec1}
In wave-control technologies across a wide range of fields, multi-band operation is a fundamental strategy that brings not merely bandwidth expansion, but also functional parallelization, improved robustness, and increased control degrees of freedom \cite{Saeidi2024,Aboagye2024,ChenY2026,TaoZ2025,JiangX2021,HuY2024}.
The simultaneous use of discretized multiple bands offers advantages such as increase in information capacity, higher sensitivity, system robustness, and reduced interference. 
Thus, the establishment of a multi-band architecture has been one of the significant technologies spreading across a wide range of fields, including communications, signal processing, sensitive sensing, and materials analysis.

The emergence of topological wave engineering has further advanced the wave manipulation technology.
By translating topological concepts originally developed in quantum condensed matter physics into classical wave systems, robust edge transport with an immunity to defect and suppression of  back-scattering has been realized in optical \cite{Khanikaev2024,Muis2026,Ozawa2019,YangY2025,Hashemi2025,GaoJ2024,WangY2023,XueH2021,Iwamoto2021}, acoustic \cite{XueH2022,Alu2025,HuangL2026,LiangX2025,Darabi2020,ZhouY2023,NiX2023,XueH2019}, electromagnetic \cite{Qian2023,TaoL2024,Thomas2026,LuT2025,Bernevig2022,WangZ2009,Gilbert2021,Haver2023}, and mechanical platforms \cite{Funayama2022, LiS2018,YuS2018,YanM2018,LiuY2025,ChenZ2025,Mousavi2015,LiG2020,WangZ2021}.
In recent years, the field of topological wave engineering has expanded beyond the stage of fundamental physics demonstration to applied engineering focusing on functionalization and implementation technologies \cite{Hata2025,WangW2024,ZhangW2025,Funayama2025,Sohn2025}.

The integration of topological band theory with multi-band architectures is a compelling direction toward practical and functional topological wave systems.
Extension of the topological wave systems to multi-band operation has been explored by coexisting fundamental and high-frequency modes and utilizing them.
Conventional high-frequency topological (CHFT) modes rely on spatially complex profiles.
The quantum valley Hall (QVH) systems offer the design advantage that the topological properties of the fundamental mode are easily inherited by those of the second topological mode, meaning that the topological fundamental mode ensures the existence of second topological mode and its topological characteristics. 
This stems from the fact that the QVH topology is primarily characterized by the sign of the Dirac mass defined near each valley (K and K'), reflecting the local nature of the band structure.
Leveraging the advantage, the QVH systems have attracted attention for multi-band topological systems with applications in nonlinear wave engineering and quantum technology \cite{ZhangZ2022,TangG2024,MaJ2021}.

On the other hand, another representative topology, namely the quantum spin Hall (QSH), is characterized by a topological invariant such as the spin Chern number defined over all occupied bands below a given band gap, reflecting its global nature in contrast to valley-dependent topology \cite{Sheng2006,Prodan2009}.

Consequently, while quantum spin Hall systems generally provide stronger symmetry-protected edge transport in the fundamental topological mode than that for QVH systems, the CHFT modes in QSH are necessary to be individually designed regardless of other topological modes and their characteristics.
Thus, the QSH-based multi-band architectures tend to become complex due to an increase in structural parameters required to design distinct topological features for each CHFT mode.
So far, the multi-band topology in QSH systems has been demonstrated through sophisticated methods such as lattice superposition and structural optimization \cite{HuangZ2022,Nanthakumar2019,ChenY2022,ChenY2024}.
However, in both platforms of QVH and QSH, scalability of multi-band topological systems based on CHFT modes is inherently constrained because the complex spatial mode profiles for the CHFT modes are concerned to induce loss of topological protection due to symmetry fragility and pseudospin hybridization \cite{XieB2020,HuangZ2022}.
As the issues argued above, a scalable design principle for arbitrary multi-band topological systems has not been achieved yet.

Here, we propose a generalized design principle of N-fold replications of a topological mode along the frequency domain without invoking CHFT modes for multi-band topological architectures.
The topological state replication has been achieved by introducing hierarchical resonators within a host structure of QSH-based sub-lattices as an internal degree of freedom.
In this scheme, we show that the same number of replicated fundamental topological modes as the number of introduced hierarchical resonators appear in the band dispersion as high-frequency topological modes with maintaining the topological characteristics for the fundamental mode.
By overcoming the design limit of multi-band topological systems with the fragile symmetry due to the complex spatial profiles of CHFT modes, we experimentally demonstrate robust and simultaneous transmission of the fundamental and replicated high-frequency topological (RHFT) modes in the linear condition.
Our results not only discover a physical concept of replication of topology in topological phase design, but also propose a paradigm expanding the functionality and designability of topological systems.

\section{Results}\label{sec2}
\subsection{General principle of topological band replication}\label{subsec1}
\begin{figure}[t!]
    \centering
    \includegraphics[width=1.0\textwidth]{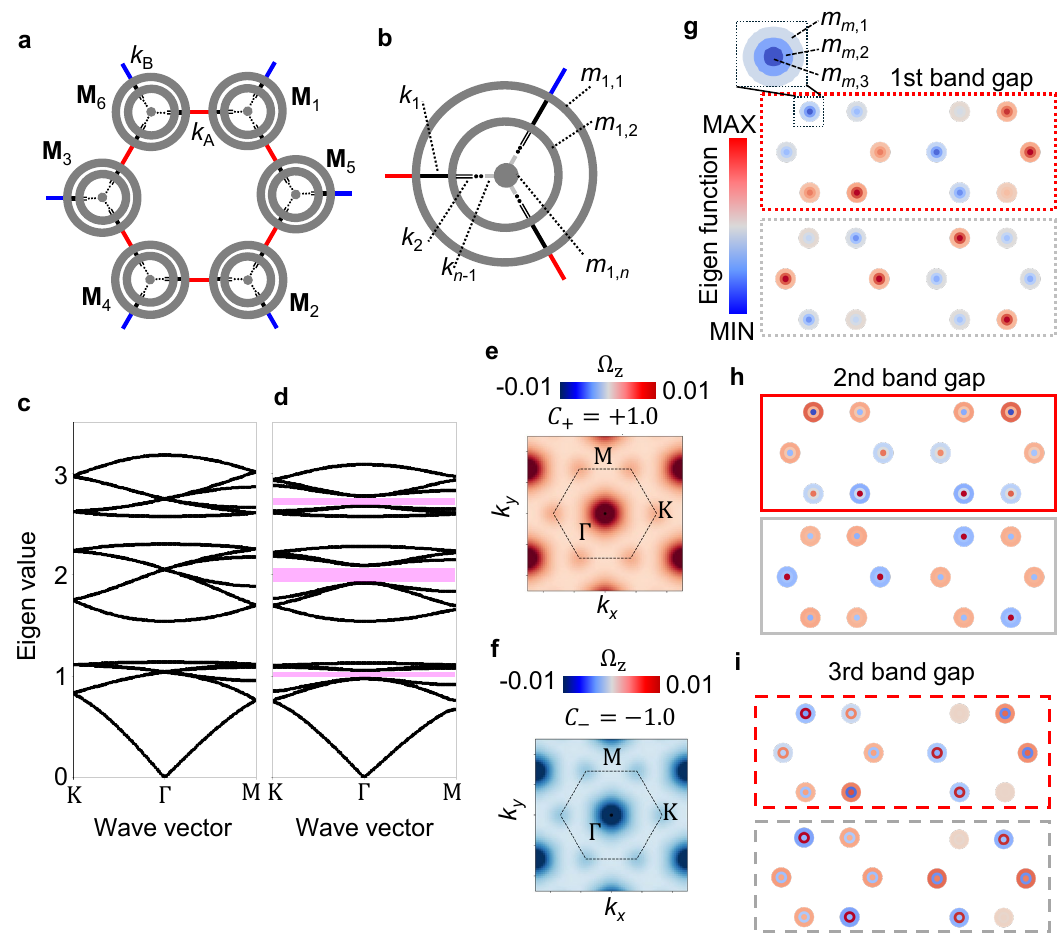}
    \caption{
        \textbf{Concept and analyses of the quantum spin Hall-based hierarchical honeycomb lattice.}
        \textbf{a} Schematic of the quantum spin Hall (QSH)-based unit cell having hierarchical resonators in each sub-lattice.
        The nearest neighboring host frameworks of the sub-lattices are connected by $k_{\mathrm{A}}$ (red line) and $k_{\mathrm{B}}$ (blue line) within the unit cell and between neighboring unit cells, respectively.
        The sub-lattices are numbered from 1 to 6.
        \textbf{b} Enlarged view of the sub-lattice labeled by numbered as 1 in (\textbf{a}).
        The hierarchical masses and connections are generalized as $m_{m,n}$ and $k_{n-1}$, respectively.
        Here, the subscripts $m$ and $n$ represent the sub-lattice number in the unit cell and hierarchical number of the components in the sub-lattice, respectively.
        \textbf{c,~d} Dispersion diagrams of the unit cell consisting of the host framework and two hierarchical inner resonators for (\textbf{c}) $k_{\mathrm{A}}=k_{\mathrm{B}}$ and (\textbf{d}) $k_{\mathrm{A}}>k_{\mathrm{B}}$.
        The shaded magenta regions in (\textbf{d}) represent the band gaps appearing by opening the band crossing at $\Gamma$ point in (\textbf{c}).
        \textbf{e,~f} Two-dimensional maps of Berry curvature $\Omega_{\mathrm{z}}$ for the first band gap in (\textbf{d}), showing (\textbf{e}) pseudo-spin-up and (\textbf{f}) pseudo-spin-down.
        $C_{\pm}$ indicates the sum of $\Omega_{\mathrm{z}}$ within the hexagonal first Brillouin zone.
        \textbf{g-h} Eigenfunction profiles in the topological unit cell at the doubly degenerated modes above (red rectangle) and below (gray rectangle) the (\textbf{g}) first, (\textbf{h}) second, (\textbf{i}) third bulk band gaps in (\textbf{d}) at $\Gamma$ point.
        In the inset of (\textbf{g}), the three colored layers represent the eigenfunction on the sub-lattice with $m=6$ consisting of host framework ($m_{m,1}$) and two hierarchical resonators ($m_{m,2}$ and $m_{m,3}$).
        }
    \label{fig:figure1}
\end{figure}

We consider a simple mass-spring model-based honeycomb lattice consisting of six sub-lattices with ring-shaped hierarchical resonators to realize topological band replication (Fig.~\ref{fig:figure1}a).
The nearest neighboring outermost-rings, which are host structures of the sub-lattices, are connected by intra-hopping $k_\mathrm{A}$ (red line) within the unit cell.
Neighboring unit cells are connected through inter-hopping $k_\mathrm{B}$ (blue line).
Each of the six sub-lattices has equal mass $\mathbf{M}_m$, where $m$ is the sub-lattice index.
Figure~\ref{fig:figure1}b indicates an enlarged schematic of the sub-lattice numbered as $m=1$.
In the general model, the sub-lattice is composed of ring-shaped hierarchical resonators with each mass of $m_{m,n}$, where $n$ is the hierarchical index from the outermost resonator, i.e., the host framework for $n=1$.
For $n\geq2$, the nearest neighboring hierarchical resonators with masses of $m_{m,n-1}$ and $m_{m,n}$ are coupled by three-coupling of $k_{n-1}$. 
Under the setup, we derive the eigenequation of the honeycomb lattice with Hamiltonian $\bm{\mathcal{H}}$ using a tight-binding method, i.e.,
\begin{align}
\label{eq:1}
\omega^2 \bm{\mathcal{U}}=\bm{\mathcal{H}}\bm{\mathcal{U}},
\end{align}
where $\omega$ is an eigenfrequency, $\bm{\mathcal{U}}$ is a vector consisting of eigenfunction vectors $\bm{U}_m$ of the six sub-lattices as $\bm{\mathcal{U}}=[\bm{U}_1\cdots \bm{U}_6]^\mathsf{T}$.
Here, the eigenfunction vector at each sub-lattice is generalized as $\bm{U}_m=[u_{m,1} \cdots u_{m,N}]^\mathsf{T}$ for $1\leq n \leq N$, where $N$ is a positive integer.
Hamiltonian $\bm{\mathcal{H}}$ is composed of mass-matrix $\bm{\mathcal{M}}$ and coupling-matrix $\bm{\mathcal{K}}(\beta)$ as
\begin{align}
\label{eq:2}
\bm{\mathcal{H}} &= \bm{\mathcal{M}}^{-1}\bm{\mathcal{K}}(\beta)\notag\\
&=
\begin{bmatrix}
\bm{M}_{1} & \cdots & \bm{O}\\
\vdots & \ddots & \vdots\\
\bm{O} & \cdots & \bm{M}_{6}
\end{bmatrix}^{-1}
\begin{bmatrix}
\bm{K} & \bm{O} & \bm{O} & -\bm{k}_\mathrm{B}e^{i\beta\cdot \mathbf{r}} & -\bm{k}_\mathrm{A} & -\bm{k}_\mathrm{A}\\
\bm{O} & \bm{K} & \bm{O} & -\bm{k}_\mathrm{A} & -\bm{k}_\mathrm{A} & -\bm{k}_\mathrm{B}e^{-i\beta\cdot \mathbf{r}}\\
\bm{O} & \bm{O} & \bm{K} &  -\bm{k}_\mathrm{A} & -\bm{k}_\mathrm{B}e^{-i\beta\cdot \mathbf{r}} & -\bm{k}_\mathrm{A}\\
-\bm{k}_\mathrm{B}e^{-i\beta\cdot \mathbf{r}} & -\bm{k}_\mathrm{A} & -\bm{k}_\mathrm{A} & \bm{K} & \bm{O} & \bm{O}\\
-\bm{k}_\mathrm{A} & -\bm{k}_\mathrm{A} & -\bm{k}_\mathrm{B}e^{i\beta\cdot \mathbf{r}} & \bm{O} & \bm{K} & \bm{O}\\
-\bm{k}_\mathrm{A} & -\bm{k}_\mathrm{B}e^{i\beta\cdot \mathbf{r}} & -\bm{k}_\mathrm{A} & \bm{O} & \bm{O} & \bm{K}
\end{bmatrix},
\end{align}
where $\bm{M}_{m}$, $\bm{K}$ and $\bm{k}_\mathrm{A(B)}$ are $N\times N$ matrices and are expressed as
\begin{align}
\label{eq:3}
\bm{M}_m=
\begin{bmatrix}
m_{m,1} & \cdots & 0\\
\vdots & \ddots & \vdots\\
0 & \cdots & m_{m,N}
\end{bmatrix},
\end{align}
\begin{align}
\label{eq:4}
\bm{K}=
\begin{bmatrix}
2k_{\mathrm{A}}+k_{\mathrm{B}}+3k_{1} & -3k_1 & \cdots & \cdots &0\\
-3k_1 & 3(k_1+k_2) &  & & \vdots\\
0 & & \ddots & & 0\\
\vdots & & & 3(k_{N-2}+k_{N-1}) & -3k_{N-1}\\
0 & \cdots & & -3k_{N-1} & 3k_{N-1}
\end{bmatrix},
\end{align}
and
\begin{align}
\label{eq:5}
\bm{k}_\mathrm{A(B)}=
\begin{bmatrix}
k_\mathrm{A(B)} & 0 & \cdots & 0\\
0 & 0 &  & \vdots\\
\vdots &  & \ddots & \vdots\\
0 & \cdots & \cdots & 0
\end{bmatrix}.
\end{align}

Hereafter, as an illustration, we analytically investigate the quantum spin Hall-based honeycomb lattice having the hierarchical resonators for $N=3$.
Under the condition, the honeycomb lattice consists of the host framework and two internal resonators.
To solve Eq.~\eqref{eq:1} for $N=3$, we set the parameters for the non-trivial topological (equal) honeycomb lattice as follows: $m_{m,1}=1.0$, $m_{m,2}=0.5$, $m_{m,3}=0.25$, $k_\mathrm{A}=0.93~(1.0)$, $k_\mathrm{B}=1.63~(1.0)$, $k_1=0.5$, and $k_2=0.25$.
Figures~\ref{fig:figure1}c and \ref{fig:figure1}d show the band diagrams of the equal and topological unit cells, respectively.
For $k_\mathrm{A}=k_\mathrm{B}$, we see three band crossing points based on quadruple degeneracy (Fig.~\ref{fig:figure1}c).
On the other hand, the asymmetric coupling for $k_\mathrm{A}=0.93$ and $k_\mathrm{B}=1.63$ opens three band gaps (shaded magenta regions) (Fig.~\ref{fig:figure1}d).
As with conventional QSH systems, there are doubly degenerate modes above and below each band gap.
Thus, we see that the number of the band gaps increases according to the number of the introduced hierarchical resonators.
To evidently show the generality of our proposal, we analytically clarify that the two and four bulk band gaps emerge in the band diagrams for $N=2$ and $4$, respectively, in section 1 of supporting information. 

The quantum spin Hall phase is characterized by the spin Chern number \cite{Sheng2006,Prodan2009}, 
\begin{align}
\label{eq:6}
C_{\mathrm{S}}&=\frac{1}{2}(C_{+}-C_{-}),\\
C_{+(-)}&=\frac{1}{2\pi}\int_{\mathrm{BZ}}\Omega\mathrm{z}_{+(-)}(\beta)\mathrm{d}^{2}\beta, 
\end{align}
which serves as the topological invariant distinguishing the topological and trivial phases.
Here, $\mathrm{BZ}$ represents the first Brilliant zone, and $\Omega \mathrm{z}_{+(-)}$ is the Berry curvature of the pseudospin-up (pseudospin-down) and calculated with occupied bands below the relevant band gap.
Figure~\ref{fig:figure1}e (\ref{fig:figure1}f) shows the 2D map of $\Omega \mathrm{z}_{+(-)}$ for the first bulk band gap in Fig.~\ref{fig:figure1}d.
The hexagon in the 2D maps represents the $\mathrm{BZ}$, and the color bar indicates an intensity of $\Omega \mathrm{z}_{+(-)}$.
Summation of the $\Omega \mathrm{z}_{+(-)}$ within $\mathrm{BZ}$ yields $C_{+(-)}=1.0(-1.0)$, resulting in $C_{\mathrm{S}}=1.0$.
This means that the first band gap hosts a non-trivial topological phase characterized by the non-zero spin Chern number.
When we evaluate the topological states based on the other band gaps in the QSH systems, spin Chern number $C_{\mathrm{S}}$ is redefined by all occupied bands below the target band gap.
In our model for $N=3$, we find that the other two high-frequency bands become topological phase with $C_{\mathrm{S}}=1.0$ similar to that for the first band.
On the other hand, we confirm that all three band gaps indicate $C_{\pm}=0$ for the honeycomb lattice with $k_\mathrm{A}=1.4$ and $k_\mathrm{B}=0.7$ ($k_\mathrm{A} > k_\mathrm{B}$), resulting in the trivial state.
The 2D maps of Berry curvature at the three band gaps for topological and trivial phases are summarized in section 2 of the supporting information. 
As argued above, interestingly, the three bands having identical topological characteristics emerge by simply increasing the hierarchical components without individually designing each band.

To further investigate the relation between the topological phases characterized by the three bulk band gaps, we visualize the profiles of the modes above and below each band gap at $\Gamma$ point for the topological honeycomb lattice as shown in Figs.~\ref{fig:figure1}g-\ref{fig:figure1}i.
In Figs.~\ref{fig:figure1}g-\ref{fig:figure1}i, the two profiles enclosed by gray (red) rectangles represent the doubly degenerate modes below (above) each band gap. 
As shown in the inset of Fig.~\ref{fig:figure1}g, the circle consisting of three color layers represents the sub-lattice with $m=6$.
The three regions having distinct colors indicate the hierarchical components $m_{m,1}$, $m_{m,2}$, and $m_{m,3}$ from the outermost layer.

First, Figure~\ref{fig:figure1}g shows that the two eigenfunction maps in the lower (upper) panel have the profiles like d-orbitals (p-orbitals) with a period of $4\pi$ ($2\pi$). 
The host framework ($m_{m,1}$) and two hierarchical resonators ($m_{m,2}$ and $m_{m,3}$) exhibit the same sign of the eigenfunction within each sub-lattice, indicating that the three components operate in-phase at the modes around the first band gap.
Secondly, Figure~\ref{fig:figure1}h (\ref{fig:figure1}i) shows the eigenfunction maps at the two doubly degenerate modes below and above the second (third) band gap similar to those in Fig.~\ref{fig:figure1}g, respectively.
In Fig.~\ref{fig:figure1}h, the host framework ($m_{m,1}$) and second layer resonator ($m_{m,2}$) show the same sign of the eigenfunction, while only the innermost resonator ($m_{m,3}$) indicates the opposite sign, i.e. out-of-phase operation.
On the other hand, for the mode profiles based on the third band gap in Fig.~\ref{fig:figure1}i, the innermost resonator ($m_{m,3}$) becomes in-phase with the host framework ($m_{m,1}$), and the second layer resonator ($m_{m,2}$) operates out-of-phase.

Interestingly, we find that the host framework maintains the p-orbital-like and d-orbital-like profiles at the modes above and below the three band gaps, respectively. 
These results show that the fundamental modes are preserved at high frequency modes without requiring complex spatial profiles in the host framework. 
This is achieved by shifting the degrees of freedom from spatial complexity to the number of hierarchical components.
Furthermore, topological state in the honeycomb lattice is determined by only the ratio between intra- and inter-coupling ($k_{\mathrm{A}}$ and $k_{\mathrm{B}}$) regardless of the parameters of hierarchical components, i.e. $m_{m,n}$ and $k_{n-1}$ ($n\geq2$).
Indeed, when the intra- and inter-coupling are equal, the change of parameters of hierarchical structures do not open any band gaps as shown in section 3 of the supporting information.

\begin{figure}[b!]
    \centering
    \includegraphics[width=1.0\textwidth]{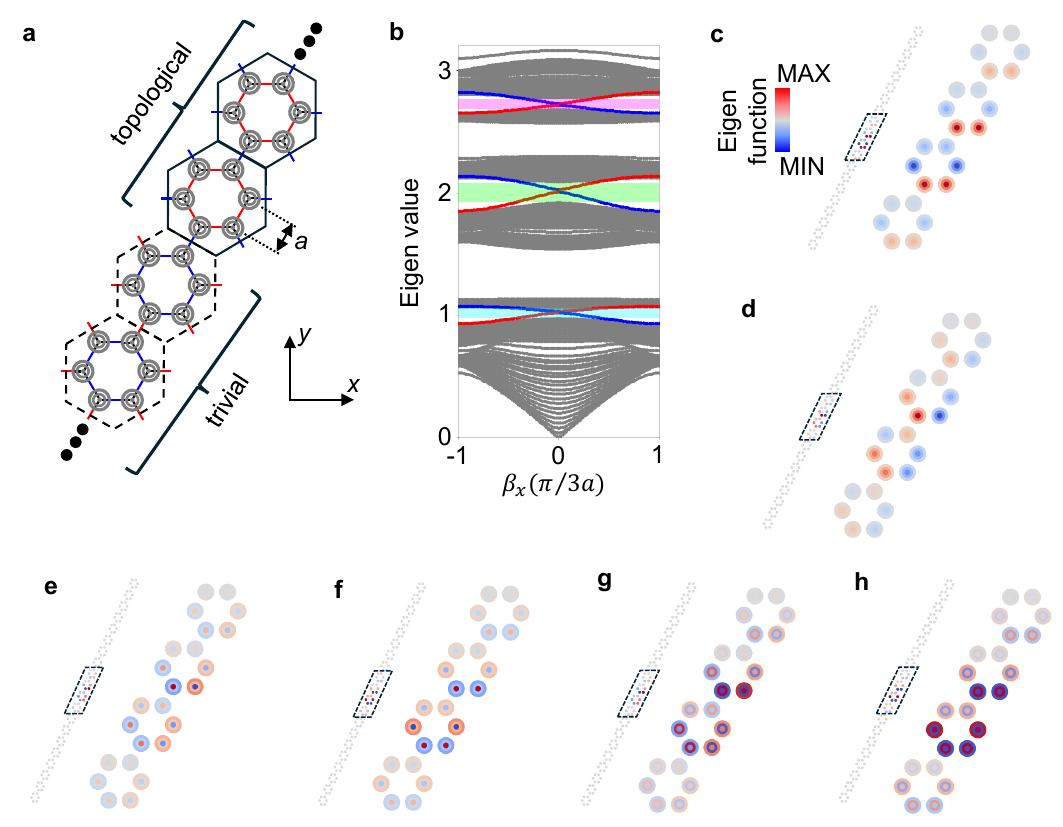}
    \caption{
        \textbf{Analyses of the supercell consisting of the hierarchical honeycomb lattice for $N=3$.}
        \textbf{a} Schematics of the supercell with the topological boundary between the quantum spin Hall (QSH)-based topological and trivial unit cells. 
        The supercell is assumed to have the infinite periodic condition along the $x$-axis.
        Along the $y$-axis, 10-topological and 10-trivial units are aligned.
        \textbf{b} Dispersion diagram of the supercell.
        Pairs of topological interface modes (red and blue lines) appear within the first (shaded cyan), second (shaded green), and third (shaded magenta) bulk band gaps.
        \textbf{c-h} Eigenfunction profiles for the interface modes for $\beta_x=0$ within the (\textbf{c}, \textbf{d}) first, (\textbf{e}, \textbf{f}) second, and (\textbf{g}, \textbf{h}) third bulk band gaps.
        The mode profiles of four units are enlarged views within the parallelogram of each overall mode profile.
        }
    \label{fig:figure2}
\end{figure}

As the next step, we investigate that the topological interface modes in the supercell composed of topological and trivial honeycomb lattices with hierarchical resonators.
The prepared supercell consists of 10-topological and 10-trivial unit cells aligned in the direction of the $y$-axis, and has the periodic boundary along the $x$-axis as shown in Fig.~\ref{fig:figure2}a.
Note that we set the structural parameters for the trivial lattice with $N=3$ as follows: $m_{m,1}=1.0$, $m_{m,2}=0.5$, $m_{m,3}=0.25$, $k_\mathrm{A}=1.4$, $k_\mathrm{B}=0.7$, $k_1=0.5$, and $k_2=0.25$.
Solving the eigenequation derived from tight-binding model with Bloch’s theorem similar to that for the unit cell, we obtain the dispersion diagram of the supercell for $N=3$ as show in Fig.~\ref{fig:figure2}b.
We see that the supercell has the band crossing topological interface modes (red and blue lines) at the first (shaded cyan), second (shaded green), and third (shaded magenta) bulk band gaps. 
Note that $a$ in the horizontal axis in Fig.~\ref{fig:figure2}b represents the center-to-center distance between the nearest neighboring sub-lattices as shown in Fig.~\ref{fig:figure2}a.

Eigenfunction profiles of the degenerated topological interface modes at $\beta_{x}=0$ in each bulk band gap are shown in Figs.~\ref{fig:figure2}c-\ref{fig:figure2}h.
Figures~\ref{fig:figure2}c and \ref{fig:figure2}d indicate the profiles for the fundamental topological interface modes in the first bulk band gap.
In addition to the localized profiles on the topological boundary, we see that the host framework and hierarchical resonators exhibit the same sign of the eigenfunctions in each sub-lattice, indicating that three resonators within each sub-lattice operate in-phase similar to that in Fig.~\ref{fig:figure1}g.
Figures~\ref{fig:figure2}e and \ref{fig:figure2}f (\ref{fig:figure2}g and \ref{fig:figure2}h) indicate the profiles at the second (third) topological interface modes in Fig.~\ref{fig:figure2}b.
In comparison with the profiles in Figs.~\ref{fig:figure2}c and \ref{fig:figure2}d, we find that the host frameworks in Figs.~\ref{fig:figure2}e and \ref{fig:figure2}f (\ref{fig:figure2}g and \ref{fig:figure2}f) consistently maintain the profiles based on the fundamental topological interface modes, while only the innermost (second-layer) hierarchical resonators have the eigenfunctions with opposite sign to those of other components in each sub-lattice for the second (third) topological interface modes.
Note that, as is clear from the proposed model, the increase in the topological interface modes is clearly different from band folding due to the expansion of lattice constant because the all hierarchical components have the same center of mass and each topological interface mode is completely discretized.
These results evidently verify that the fundamental topological interface modes, which are characterized by the host framework, are replicated as the high-frequency topological interface modes by extending the number of hierarchical resonators to degree of freedom, i.e. the RHFT modes.
The model of hierarchical resonators, which are feasible design in diverse wave systems, provides a general design principle for topological systems with arbitrary multi-bands.
Furthermore, the RHFT modes fundamentally solve long-standing issues of the multi-band topology such as an absence of consistent design guidelines and the fragile symmetry to perturbations by the complex spatial profiles at the CHFT modes.

\subsection{Wave guiding demonstration of the replicated higher-frequency topological interface mode}\label{subsec2}
\begin{figure}[t!]
    \centering
    \includegraphics[width=1.0\textwidth]{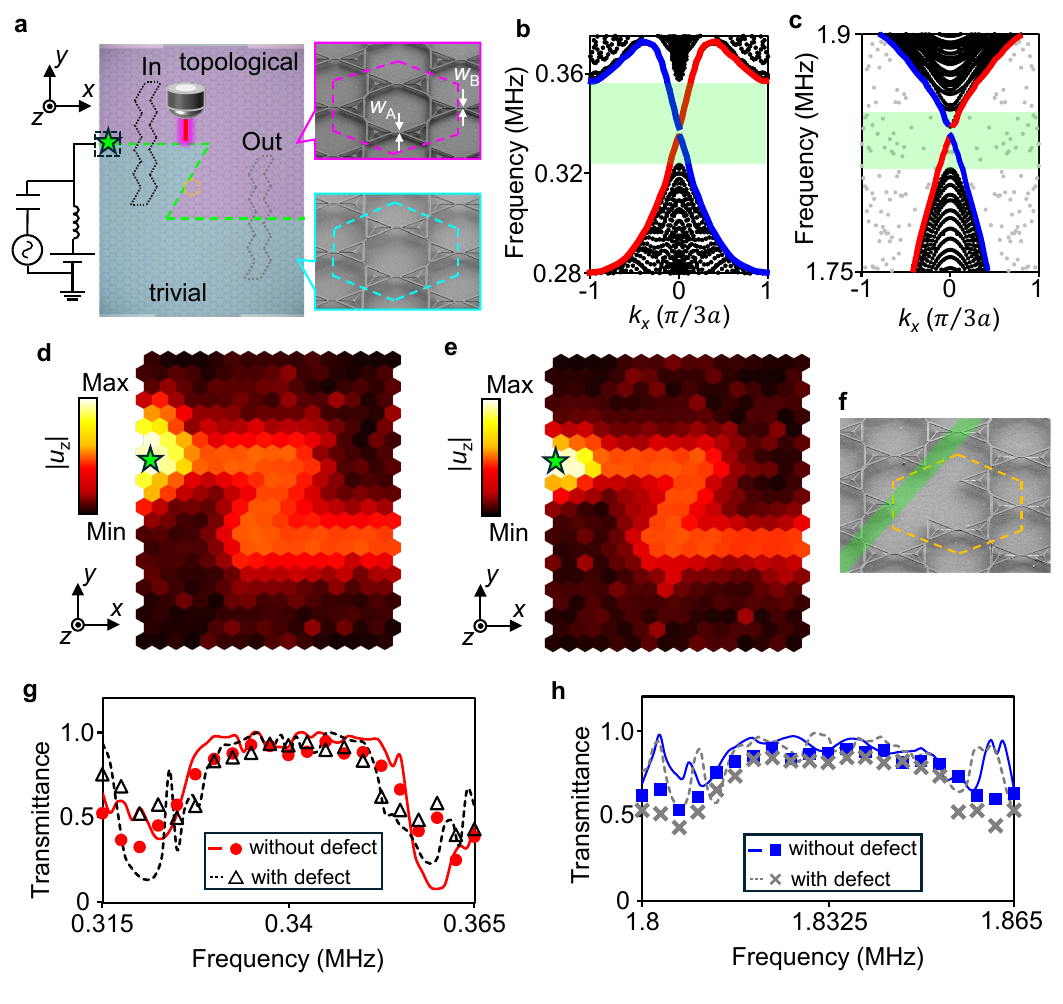}
    \caption{
        \textbf{Dual band topological waveguide.}
        \textbf{a} Top-view of the micro-electromechanical structure having a Z-shaped topological interface (broken green line).
        The areas above (shaded magenta) and below (shaded cyan) the interface are composed of topological and trivial unit cells as shown in the upper and lower scanning electron microscope (SEM) images, respectively.
        Topological and trivial phases are designed by the ratio of the intra- and inter-beam widths $w_{\mathrm{A}}$ and $w_{\mathrm{B}}$.
        \textbf{b,~c} Numerical dispersion diagrams around the (\textbf{b}) first and (\textbf{c}) second bulk band gap (shaded green region) for the supercell having the topological boundary.
        The supercell consists of the unit cells with $N=2$. 
        The blue and red lines, black symbols, and gray symbols represent the topological interface modes,  transverse bulk modes, and longitudinal bulk modes, respectively. 
        \textbf{d,~e} Experimentally obtained displacement maps on the microelectromechanical structure for excitation at (\textbf{d}) 0.345 MHz and (\textbf{e}) 1.828 MHz.
        The green star shows the excitation point similar to that in (\textbf{a}).
        \textbf{f} SEM image of a unit cell lacking two sub-lattices on the topological interface (shaded green region).
        \textbf{g,~h} Frequency spectra of transfer efficiencies around (\textbf{g}) the first and (\textbf{h}) second bulk band gaps.
        The lines and symbols represent numerical and experimental results.
    }
    \label{fig:figure3}
\end{figure}

Here, we experimentally demonstrate dual-topological wave guiding based on the fundamental and RHFT modes using a versatile silicon-based microelectromechanical system (MEMS) platform.
Scanning electron microscope images in Fig.~\ref{fig:figure3}a show the prepared silicon membranes of topological (upper image) and trivial (lower image) unit cells in the 2D structure having a Z-shaped topological boundary (broken green line in the photo image).
Both unit cells consist of triangular host frameworks and inner resonators corresponding to the most basic structure in Fig.~\ref{fig:figure1}a, i.e. $N=2$.
The triangular host framework and inner resonator are connected by the three beams with equal width and length, and the topological phases of the unit cells are designed by the ratio between silicon beam widths of the intra- ($w_\mathrm{A}$) and inter-connections ($w_\mathrm{B}$).
For the topological and trivial phases, the ratios of beam widths are designed to be $w_\mathrm{B}/w_\mathrm{A}<1$ and $w_\mathrm{B}/w_\mathrm{A}>1$, respectively.

As the prior confirmation for the device fabrication, we calculate the dispersion diagram of the supercell with the topological interface consisting of the topological and trivial unit cells.
Figures~\ref{fig:figure3}b and \ref{fig:figure3}c show the dispersion diagrams corresponding to the frequency regions around the fundamental and replicated second topological bulk band gaps (shaded green regions), respectively.
We find that the pairs of topological interface modes (red and blue lines) appear within both bulk band gaps.
The details and results of numerical simulations of topological and trivial unit cells under infinite periodic condition are shown in section 4 of the supporting information, where we can see that the triangular host framework consistently maintains p-orbital ($2\pi$) and d-orbital ($4\pi$) like displacement profiles, while the internal resonator operates in-phase and out-of-phase with the host structure at the fundamental and second bulk band gaps, respectively.

Based on the numerical models, the MEMS membrane-based topological waveguide is fabricated using a silicon-on-insulator (SOI) substrate composed of thin silicon, silicon dioxide, and supporting silicon layers with the thicknesses of 1 \textmu m, 10 \textmu m, and 470 \textmu m, respectively.
After device fabrication, we experimentally measure the $z$-axis displacement, $u_z$, of each unit cell in the 2D structure by scanning laser doppler vibrometer as shown in Fig.~\ref{fig:figure3}a.
The green star represents the excitation point, where a few unit cells around the left edge of the waveguide are electrostatically excited from the electrode formed on the supporting silicon layer.
The fabrication process and measurement setup are similar to those in our previous studies \cite{Funayama2022,Funayama2024}, and the details of specific conditions are shown in Method section. 

Figures~\ref{fig:figure3}d and \ref{fig:figure3}e show the experimentally obtained 2D maps of $|u_z|$ at 0.345 MHz and 1.828 MHz corresponding to the fundamental and replicated second topological interface modes, respectively.
We see that the elastic waves propagate from the left to right edges of the identical Z-shaped waveguide at both frequencies.
The wave routing at the sharp corner without outstanding dissipation is one of the evidences of topological protection.
As one more well-known characteristic of topological protection, we verify the immunity to defects for wave propagation.
A defect is introduced to a topological unit cell highlighted by the orange hexagon in Fig.~\ref{fig:figure3}a .
Figure~\ref{fig:figure3}f shows an enlarged view of the  unit cell in the absence of two sub-lattices, where the shaded green region represents the topological boundary.
We quantitatively evaluate the topologically protected wave propagation by transmittance of the waveguide.
The transmittance is calculated by the ratio between the sum of $|u_z|^2$ in the input and output ports, which are expressed as In and Out in Fig.~\ref{fig:figure3}a.
Figures~\ref{fig:figure3}g and \ref{fig:figure3}h show the frequency spectra of the transmittance around the first and second bulk band gaps.
In both band gaps, the results in simulations (lines) and measurements (symbols) indicate highly effective wave transmission regardless of existence of the defect.
Therefore, we have verified that the replicated topological mode shows the topological protection similar to the fundamental topological interface mode.

A key function of multi-band waveguides is the ability to transmit multiple frequency channels in parallel with suppressed cross-talk under linear operation.
In topological wave systems, such multi-channel operation is further enhanced because the topological interface mode within each bulk band gap works as a channel exhibiting suppression of  backscattering and immunity to defects.
However, multi-band topological systems based on the CHFT modes are inherently susceptible to perturbation-induced symmetry breaking and pseudospin hybridization due to the complex spatial profiles.
Here, we compare the topological protection for the RHFT and CHFT modes through immunity to defect, and demonstrate practical multi-channel capability using the RHFT mode.

\begin{figure}[b!]
    \centering
    \includegraphics[width=1.0\textwidth]{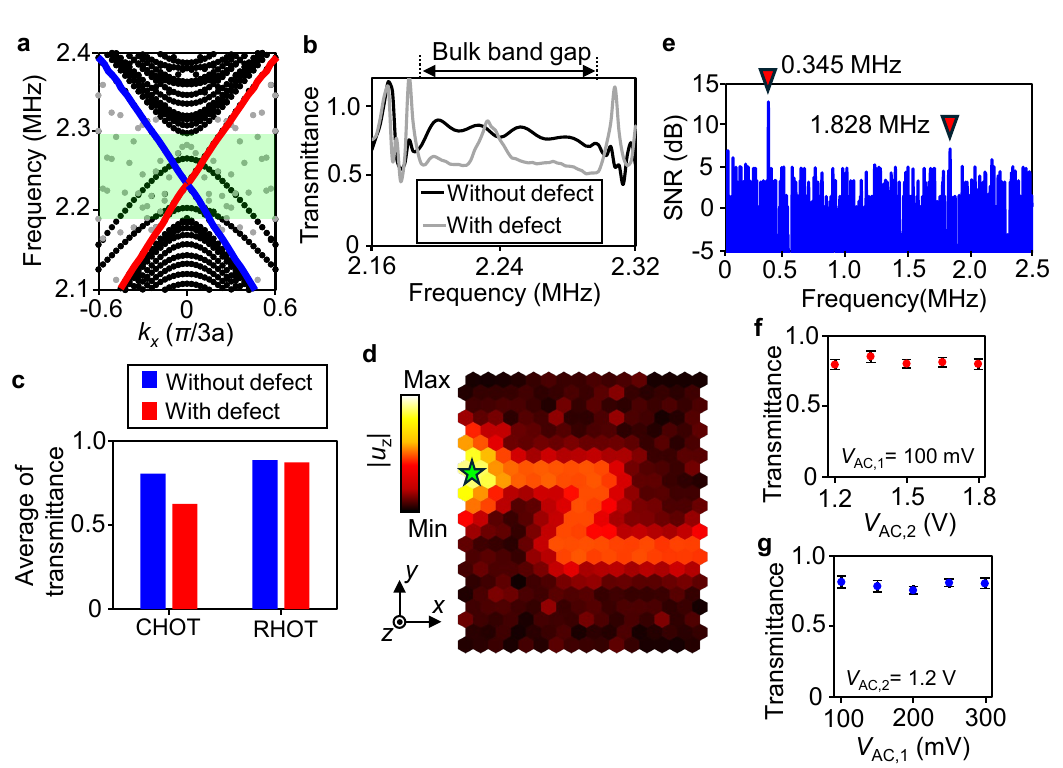}
    \caption{
        \textbf{Robustness of the replicated topological interface mode.}
        \textbf{a} Numerical dispersion diagrams around the conventional high-frequency topological interface modes of the same supercell with that for Figs.~\ref{fig:figure3}b and \ref{fig:figure3}c.
        The shaded green region represents the bulk band gap.
        The red and blue lines, black symbols, and gray symbols represent the topological interface modes, bulk and trivial edge modes, and longitudinal modes.
        \textbf{b} Numerical frequency spectra of transfer efficiencies around the conventional high-frequency bulk band gap (shaded green region in (\textbf{a})).
        \textbf{c} Averaged transfer efficiencies of the CHFT and RHFT modes across each bulk band gap.
        The blue and red bars represent the averaged efficiencies of the Z-shaped waveguides without and with the defect, respectively.
        \textbf{d} Experimentally obtained displacement map on the 2D microelectromechanical structure for simultaneous excitation of 0.345 MHz (fundamental topological mode) and 1.828 MHz (RHFT mode).
        The green star represents the excitation point.
        \textbf{e} Power spectrum density calculated from the $u_z^2$ at the output port in Fig.~\ref{fig:figure3}a.
        \textbf{f,~g} Transfer efficiencies of (\textbf{f}) the fundamental topological mode and (\textbf{g}) RHFT mode as a function of AC voltage amplitudes $V_{\mathrm{AC,2}}$ (1.828 MHz) and $V_{\mathrm{AC,1}}$ (0.345 MHz), respectively.
        Error bars indicate standard deviation of the results of six-repeated measurements.
    }
    \label{fig:figure4}
\end{figure}

At first, we compare the immunity to the defect of the CHFT and RHFT modes through the transmission efficiencies.
In the same supercell with that for Figs.~\ref{fig:figure3}b and \ref{fig:figure3}c, the CHFT modes with complex spatial profiles (red and blue lines) also appear as shown in Fig.~\ref{fig:figure4}a (see section 5 of supporting information for the details of the numerical simulation for the CHFT modes).
The gray symbols and shaded green region represent the longitudinal wave and bulk band gap, respectively.
Note that the black symbols represent the bulk modes and the trivial edge modes emerging at the edge of the finite structure.
Figure~\ref{fig:figure4}b shows the frequency spectra of transfer efficiencies similar to those in Figs.~\ref{fig:figure3}g and \ref{fig:figure3}h, around the conventional second bulk band gap.
For the Z-shaped waveguide without the defect, the CHFT modes exhibit highly efficient transmission (black line).
On the other hand, the efficiency clearly decreases when the waveguide has a defect (gray line).
To quantitatively compare the topological protection, we show the average transmittance for the CHFT and RHFT modes without (blue bar) and with (red bar) the defect as shown in Fig.~\ref{fig:figure4}c.
Each average value is calculated by the transmittance within all range of each bulk band gap.
Both topological interface modes achieve high-efficiency exceeding 80\% for the absence of defect.
While the RHFT modes maintain the high transmittance in the waveguide having the defect, the CHFT modes show drastic reduction of the transmittance to 62\%.
According to the results, the proposed RHFT modes have an excellent potential to realize the multi-band topological architectures with robustness to perturbations by avoiding the complex spatial profiles. 

Finally, we experimentally demonstrate simultaneous propagation of the fundamental topological mode of 0.345 MHz and RHFT mode of 1.828 MHz.
Here, we adjust the amplitudes of AC voltages of 0.345 MHz ($V_{\mathrm{AC,1}}$) and 1.828 MHz ($V_{\mathrm{AC,2}}$) to $V_{\mathrm{AC,1}}=100$ mV and $V_{\mathrm{AC,2}}=1.2 V$, respectively, to operate the waveguide in the linear condition.
Exciting the waveguide by a DC voltage of 40 V and the AC voltages argued above, we experimentally obtain the 2D map of $|u_z|$ for the simultaneous transmission of the two frequencies as shown in Fig.~\ref{fig:figure4}d.
We see that the excited elastic waves dominantly propagate along the Z-shaped waveguide.
Figure~\ref{fig:figure4}e shows the power spectrum density obtained at the output port shown in Fig.~\ref{fig:figure3}a.
In the measurement, we directly measure the wave forms without lock-in-amplifier to obtain the signal for all frequency components.
The result in Fig.~\ref{fig:figure4}e indicates the average of the integrated values from 20 times measurements.
We see the two distinct peaks corresponding to the excitation frequencies (0.345 MHz and 1.828 MHz) without significant peaks.
No existence of any sum or difference frequencies indicates that the waveguide is primarily driven under the linear condition.
Under the linear operation, we investigate the incoherent characteristic between the fundamental topological mode and RHFT mode.
Figure~\ref{fig:figure4}f (\ref{fig:figure4}g) shows the transmittance of the topological interface mode at 0.345 MHz (1.828 MHz) as a function of $V_{\mathrm{AC,2}}$ ($V_{\mathrm{AC,1}}$).
In Fig.~\ref{fig:figure4}f (\ref{fig:figure4}g), $V_{\mathrm{AC,1}}$ ($V_{\mathrm{AC,2}}$) is fixed to 100 mV (1.2 V). 
We find that one topological interface mode exhibits a constant transfer efficiency without being affected by the change in the amplitude of the other interface mode.
The constant transfer efficiencies in Figs.~\ref{fig:figure4}f and \ref{fig:figure4}g indicate the suppression of cross-talk between the fundamental topological mode and RHFT mode in the topological waveguide under the linear operation.
These results show that the topological waveguide has ability to robustly transmit multiple frequency signals in parallel with suppression of cross-talk between the distinct frequency bands.

\section{Discussion}\label{sec3}
As described above, we have proposed the design principle of topological replication emerging the multiple topological interface modes by hierarchical components in each sub-lattice.
Since the proposed mechanism based on multiplexing sub-lattices can be described within a general coupled-mode (or tight-binding) framework, it can represent a wide range of wave systems. 
Therefore, the design principle is applicable to other wave phenomena, such as photonic, acoustic, and electromagnetic platforms, by replacing the physical parameters.
Utilizing a universal microelectromechanical structure as one of the wave platforms, we experimentally demonstrated a dual-band topological waveguide based on the RHFT mode.
In this section, we discuss the limitations and potential for expansion of this design principle in real devices.

Even with this simple and consistent design methodology, the difficulty of designing practical devices increases as the number of RHFT modes increases.
Accordingly, design constraints such as fixed structural and material parameters are possible to prevent us from designing the practical configurations having the RHFT modes.
Indeed, the platform with uniform-thickness (1 \textmu m) in this study is difficult to adjust the parameters for the manufacturable structure of $N=3$ by manual parameter tuning. 
On the other hand, when we release the thickness as a variable parameter, the hierarchical resonators of $N=3$ is designed more easily.
Here, we show an example of the silicon-based topological unit cell of $N=3$ with manufacturable structural parameters numerically.

\begin{figure}[t!]
    \centering
    \includegraphics[width=1.0\textwidth]{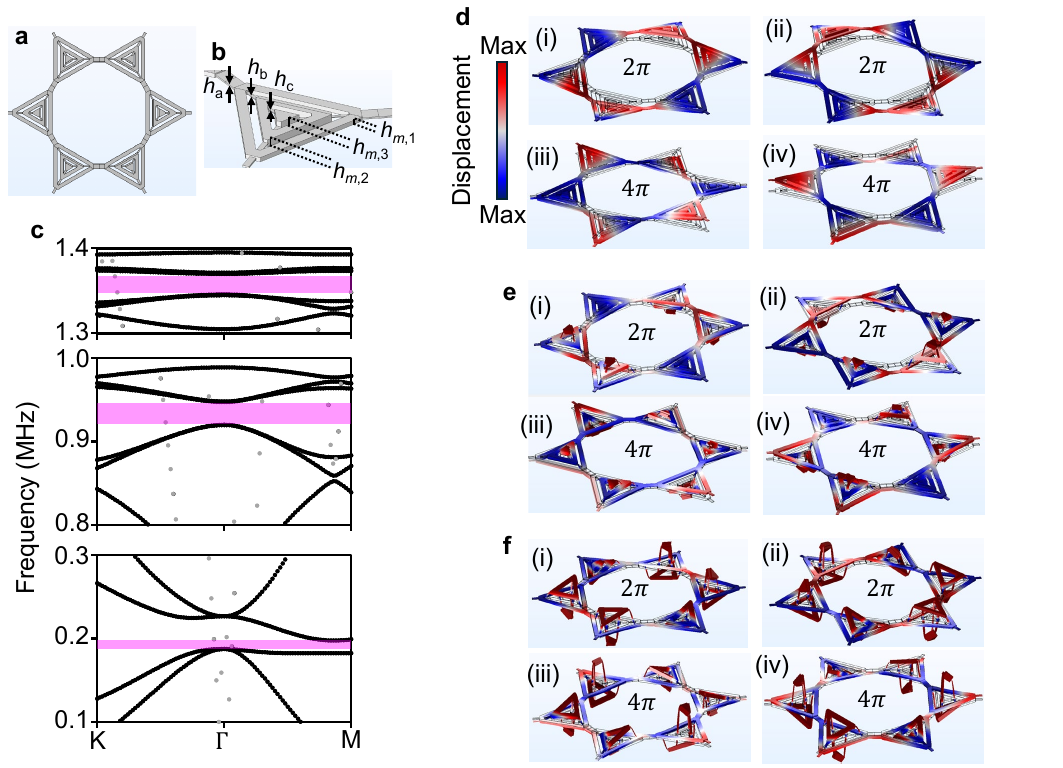}
    \caption{
        \textbf{Numerical design and analyses for a manufacturable topological unit cell with $N=3$.}
        \textbf{a} Top-overview of the QSH-based topological unit cell with $N=3$ consisting of the components of different thicknesses.
        \textbf{b} The enlarged view of a sub-lattice in the unit cell in (\textbf{a}).
        $h_{m,1}$, $h_{m,2}$, and $h_{m,3}$ indicate the thicknesses of the host framework, middle resonator, and innermost resonator, respectively.
        $h_\mathrm{a}$, $h_\mathrm{b}$, and $h_\mathrm{c}$ represent the thicknesses of the intra- and inter-beams, the beam between the host framework and middle resonator, and the beam between the middle and innermost resonators, respectively.
        \textbf{c} Dispersion diagrams of the topological unit cell in (\textbf{a}). 
        The bottom, middle, and top panels indicate the dispersion diagrams around the fundamental, second replicated, and third replicated bulk band gaps (shaded magenta regions), respectively.
        \textbf{d-f} Displacement profiles of the degenerated modes (i and ii) above and (iii and iv) below each bulk band gap at $\Gamma$ point.
        The modes classified as (i and ii) and (iii and iv) show the p-orbital ($2\pi$) and d-orbital ($4\pi$) like profiles.
    }
    \label{fig:figure5}
\end{figure}

Figure~\ref{fig:figure5}a shows one of examples of the topological unit cell with $N=3$.
The enlarged view of the unit cell as shown in Fig.~\ref{fig:figure5}b indicates that the sub-lattice consists of components with different thicknesses.
The host framework, middle resonator, and innermost resonator have the thicknesses of $h_{m,1}=1.6$ \textmu m, $h_{m,2}=2.6$ \textmu m, and $h_{m,3}=4.0$ \textmu m, respectively.
Additionally, the intra- and inter-beams, the beam between host framework and middle resonator, and the beam between middle and innermost resonators are designed to be the thicknesses of $h_\mathrm{a}=0.6$, $h_\mathrm{b}=0.24$, and $h_\mathrm{c}=0.12$ \textmu m, respectively.
Other structural parameters are shown in section 6 of supporting information.
Calculating the dispersion diagram of the unit cell, we find an emergence of three bulk band gaps as shown in Fig.~\ref{fig:figure5}c.
In the dispersion relation, the two high-frequency modes (middle and top panels) correspond to the RHFT modes.
The doubly degenerate modes above and below each bulk band gap at $\Gamma$ point indicate p-orbitals ((i) and (ii) in Figs.~\ref{fig:figure5}d-\ref{fig:figure5}f) and d-orbitals ((iii) and (iv) in Figs.~\ref{fig:figure5}d-\ref{fig:figure5}f), respectively, resulting in the topological phase as designed.
As discussed in the previous sections, the host framework maintains similar profiles across modes of the same classification from (i) to (iv) in Figs.~\ref{fig:figure5}d–\ref{fig:figure5}f, while the two inner resonators exhibit different operation states.
Through the discussion above, we validated versatility of our design principle.
Note that this design method is not limited to quantum spin Hall systems, and enables mode replication in other systems, e.g. quantum valley Hall systems.
Additionally, it would be interesting to apply the design scheme to other wave phenomena such as photonic, acoustic, and electromagnetic systems, and to explore the realization of a multi-band system with more channels under design constraints by combining the conventional excellent design methodologies such as topology optimization.

In summary, we have provided the design principle for scalable multi-band via the replication of topological fundamental mode.
The generalized design guideline has enabled us to arbitrarily replicate a fundamental topological state to the higher frequency side by the hierarchical degree of freedom of the sub-lattice in the honeycomb structure.
In both theoretical and numerical investigations by using QSH systems, the RHFT modes have shown the similar spatial profiles to those for the fundamental topological interface modes, which have been known to exhibit p-orbitals and d-orbitals like profiles in classical-wave platforms.
As an experimental demonstration, we have fabricated the dual-band topological waveguide based on the topological replication, and evaluated that the RHFT mode exhibits better immunity to defects than the CHFT mode.
We have further shown the simultaneous propagation of the fundamental topological mode and RHFT mode while suppressing cross-talk to each other under linear operation.
Our concept and results in this study open the way to the multifunctional topological systems as an implementable technology in the field of applied engineering such as next-generation communications and energy transmission infrastructure.

\section{Methods}\label{sec4}
\subsection{Numerical simulation}
We built numerical models in COMSOL Multiphysics 6.4 with MEMS module.
The dispersion diagram of the unit cell is calculated by setting the Floquet-periodic boundary condition at the cross sections of all inter-beams.
For the preparation of the supercell, 10-topological and 10-trivial unit cells are aligned perpendicular to the topological boundary.
We set the Floquet-periodic boundary condition at the cross sections of the inter-beams similar to those for the unit cell.
The details of the structural parameters for numerical simulation of the topological and trivial unit cells are shown in section 4 of supporting information. 
The numerical model having the Z-shaped waveguide consists of 16$\times$12 unit cells mimicking the experimental structure.
We set a low reflection boundary condition to all edges of the numerical models.
To replicate the experimental excitation condition, we excite the 4 unit cells at the left end of the waveguides.

\subsection{Device fabrication}
We manufactured the silicon-based honeycomb lattices consisting of the hierarchical sub-lattices on a SOI substrate by using the semiconductor processing technology.
As the first step, we spin-coated HMDS and resist (ZEP520A, Zeon Corporation) on the silicon top layer at 5000 rpm and 6000 rpm, respectively, then we pre-baked the SOI substrate at 170℃ for 5 mins.
We described the 2D pattern including the Z-shaped waveguide on the top silicon layer by electron beam (EB) lithography (JEOL JBX-6300FS).
After developing and rinsing the the substrate in o-xylene and isopropyl alcohol, respectively, the silicon top layer was etched to form the periodic honeycomb lattices by the Bosh process (MUC-21 ASE-Pegasus, Sumitomo Precision Products).
For the next step, we spin-coated HMDS and photo resist (THMR-iP3100MM, Tokyo Ohka Kogyo Co., Ltd.) at 5000 rpm.
After pre-baking for 90 seconds at 90 $^{\circ}C$, we formed the pattern of the excitation electrode on the bottom silicon layer by the photolithography (EVG620, EVG).
After the development and rinse by TMAH and deionized water, respectively,  the bottom silicon layer around the excitation electrode was also etched by the Bosh process to insulate the electrode structure from the silicon bottom layer.
Then, we removed the silicon dioxide layer beneath the periodic structure in hydrofluoric acid.
Finally, we obtained the silicon membrane consisting of the periodic honeycomb lattices by supercritical drying.

\subsection{Measurement setup}
The fabricated devices were measured in a high-vacuum condition ($\leq5.0\times10^{-5}$ Pa) to suppress the air damping.
One manual probe was contacted with the top silicon layer having the silicon membrane.
Another probe contacted with the bottom silicon layer from the top surface of the substrate through the contact hole formed by etching the top silicon and silicon dioxide layers.
Applying electric signals between the two probes, we excited the silicon membrane by the electrostatic force.
The offset DC bias and sinusoidal AC signal for the excitation were generated by a functional generator (RIGOL, DG972) and stabilized power supply (Kikusui Electronics Corporation, PMX110-0.6A), and combined with a bias tee (Tektronix Keithley Instruments, PSPL5530B).
The DC voltage was consistently 40 V in this study.
For the measurement of single-signal transmission, we fixed the AC signal amplitude with 0.1 V and 1.2 V to excite the topological interface modes of 0.345 MHz and 1.828 MHz, respectively. 
The electrostatic force $F(t)$ for the excitation is described as the following equation,
\begin{align}
\label{eq:8}
F(t)=\frac{1}{2}\frac{dC}{dz}V(t)^2, 
\end{align}
where $C$ is the effective capacitance between the silicon membrane and excitation electrode, $z$ is the distance between them, and $V(t)$ is the applied electric signal.
In our device, the maximum displacement on the waveguide was the order of $10^{-8}$ m by the numerical estimation, which was much smaller than the thickness of the silicon dioxide layer (10 \textmu m), thereby, $dC/dz$ can approximate a constant parameter.
As another significant parameter, $V(t)$ is expressed as $V(t)=V_{\mathrm{DC}}(t)+V_{\mathrm{AC}}\cos{(\omega t)}$.
Thus, the electrostatic force includes three components, i.e., 
\begin{align}
\label{eq:9}
F(t)=\frac{1}{2}\frac{dC}{dz}\left((V_\mathrm{DC}^2+\frac{1}{2}V_\mathrm{AC}^2) + 2V_\mathrm{AC}V_\mathrm{DC}\cos{(\omega t)}+\frac{1}{2}V_\mathrm{AC}^2\cos{(2\omega t)}\right)
.
\end{align}
This equation shows that we can excite the frequency of $\omega$ dominantly when $V_{\mathrm{DC}}$ is much larger than $V_{\mathrm{AC}}$.
As preparation of the measurements, we confirmed that the double-frequency ($2\omega$) components were negligibly small for the values of applied $V_{\mathrm{DC}}$ and $V_{\mathrm{AC}}$.

For the dual signal transfer, we simultaneously applied $V_{\mathrm{DC}}=40$ V, $V_{\mathrm{AC,1}}=0.1$ V for 0.345 MHz, and $V_{\mathrm{AC,2}}=1.2$ V for 1.828 MHz.
In this measurements, we directly input the signal from the pre-amplifier into the oscilloscope without lock-in amplifier when we obtained the frequency spectra.
We calculated the cumulative average of frequency spectra from 20 measurements for Fig.~\ref{fig:figure4}e.

When we evaluated the cross-talk between the topological interface modes within distinct bulk band gaps, we measured transfer efficiencies of the mode at 0.345 (1.828) MHz with varying $V_{\mathrm{AC,2}}$ ($V_{\mathrm{AC,1}}$) from 1.2 V (0.1 V) to 1.8 V (0.3 V) while $V_{\mathrm{AC,1}}$ ($V_{\mathrm{AC,2}}$) was fixed to 0.1 V (1.2 V).
To evidently show the suppression of cross-talk between the two topological interface modes under the linear operation, we plotted the average values and error bars of standard deviation calculated by the results of six repeated measurements (Figs.~\ref{fig:figure4}f and \ref{fig:figure4}g).

\bmhead{Supplementary information}
Supporting Information is available free of charge.
Details of the schematics and dispersion diagrams of the hierarchical structures for $N=2,~4$,
the 2D Berry phases of each band for topological and trivial phases, 
the dependence of dispersion diagram on structural parameters,
the simulation model and numerical results for device fabrication,
the numerical results of conventional higher-frequency topological interface with complex spatial profile, and
the numerical model and results for the phononic crystal for $N=3$.

\bmhead{Acknowledgements}
A part of this work was supported by Nagoya University microstructural characterization platform as a program of ‘Advanced Research Infrastructure for Materials and Nanotechnology in Japan (ARIM)’ of the Ministry of Education, Culture, Sports, Science and Technology (MEXT), Japan.

\bmhead{Data availability}
All data for validating this paper are provided as a Source Data file. 
Additional data supporting this paper are available from the corresponding author upon request.

\bmhead{Author contributions}
K.F. performed simulations, fabrications, experiments, and analysis. K.Y. supported the reinforcement of theoretical aspects. H.I. contributed to discussions on theoretical and numerical analysis and supervised this project. All authors contributed to discussions and manuscript preparation.

\bmhead{Competing interests}
The authors declare no competing interests.

\bibliography{sn-bibliography}

\end{document}


\clearpage

\section{Section 1: Dispersion diagrams for $N=2$ and $4$}
\begin{figure}[h!]
	\centering
	\includegraphics[width=1.0\textwidth]{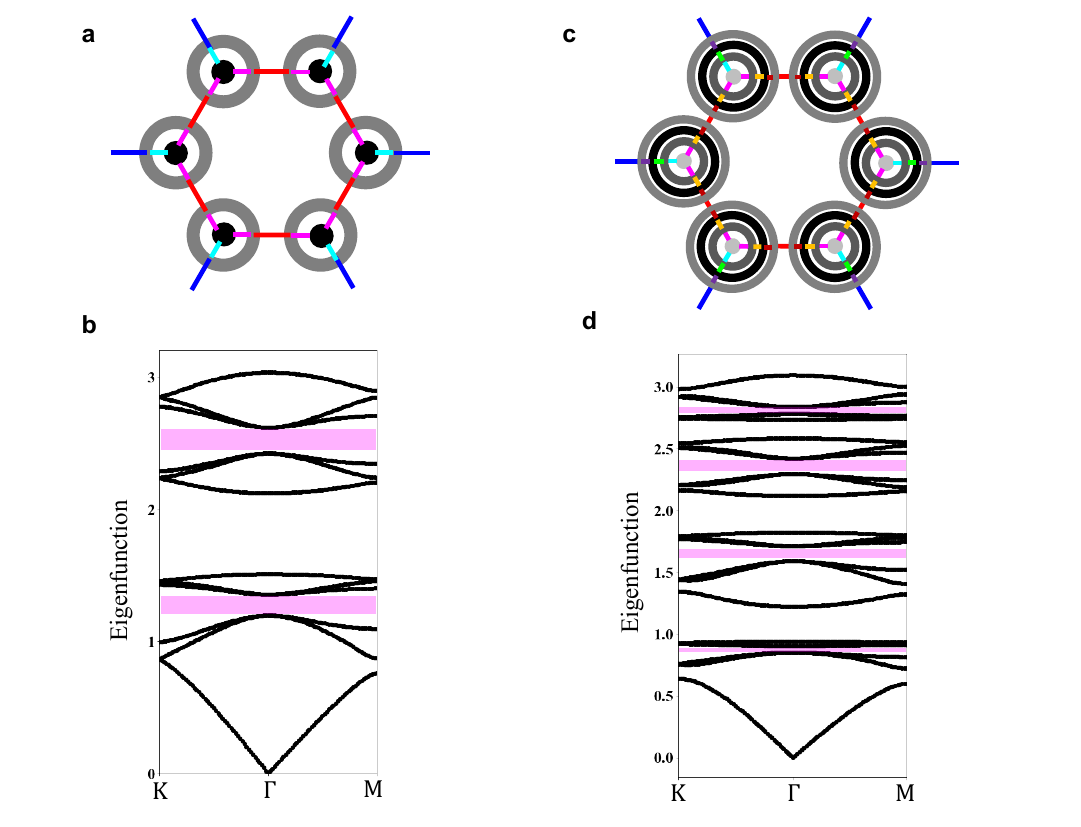}
	\caption{
        \textbf{a} Schematic of the topological unit cell with $N=2$.
        \textbf{b} Dispersion diagram of the unit cell in (\textbf{a}).
        The shaded magenta regions represent the complete band gaps. 
        \textbf{c} Schematic of the topological unit cell with $N=4$.
        \textbf{d} Dispersion diagram of the unit cell in (\textbf{c}).
        The shaded magenta regions represent the complete band gaps. 
        }
	\label{fig:figureS1}
\end{figure}

Here, we investigate the relation between the number of hierarchical components and the number of bulk band gaps.
Figure~\ref{fig:figureS1}a shows the schematic of the unit cell with $N=2$.
The parameters in the unit cell are set as follows: $m_{\mathrm{m,1}}=1.0$, $m_{\mathrm{m,2}}=0.5$, $k_{\mathrm{A}}=0.93$, $k_{\mathrm{B}}=1.63$, and $k_{\mathrm{1}}=0.5$.
Based on the parameters, we obtain the dispersion diagram as shown in Fig.~\ref{fig:figureS1}b when we solve the eigenvalue equation derived as shown in Eq.~(1) in the main manuscript.
We see that the dispersion diagram has the two complete band gaps (shaded magenta regions), where no eigenmodes exist for any wave vector in the Brillouin zone.

Similar to the case for $N=2$, we prepare the model of the honeycomb lattice with $N=4$ as shown in Fig.~\ref{fig:figureS1}c.
In addition to the parameters for the unit cell with $N=2$, the unit cell with $N=4$ is designed using the following additional parameters: $m_{\mathrm{m,3}}=0.25$, $m_{\mathrm{m,4}}=0.125$, $k_2=0.25$, and $k_3=0.125$.
Figure~\ref{fig:figureS1}d shows the dispersion diagram of the unit cell with $N=4$.
We see that the four complete band gaps (shaded magenta regions) emerge by increasing the number of the hierarchical components.

As the consistent results for $N=2$, 3, and 4, we show that the number of the complete bulk band gaps corresponds to the number of the hierarchical components in the sub-lattice.

\clearpage

\section{Section 2: Berry phase maps and spin Chern number for honeycomb lattice consisting of hierarchical components}

\begin{figure}[h!]
	\centering
	\includegraphics[width=1.0\textwidth]{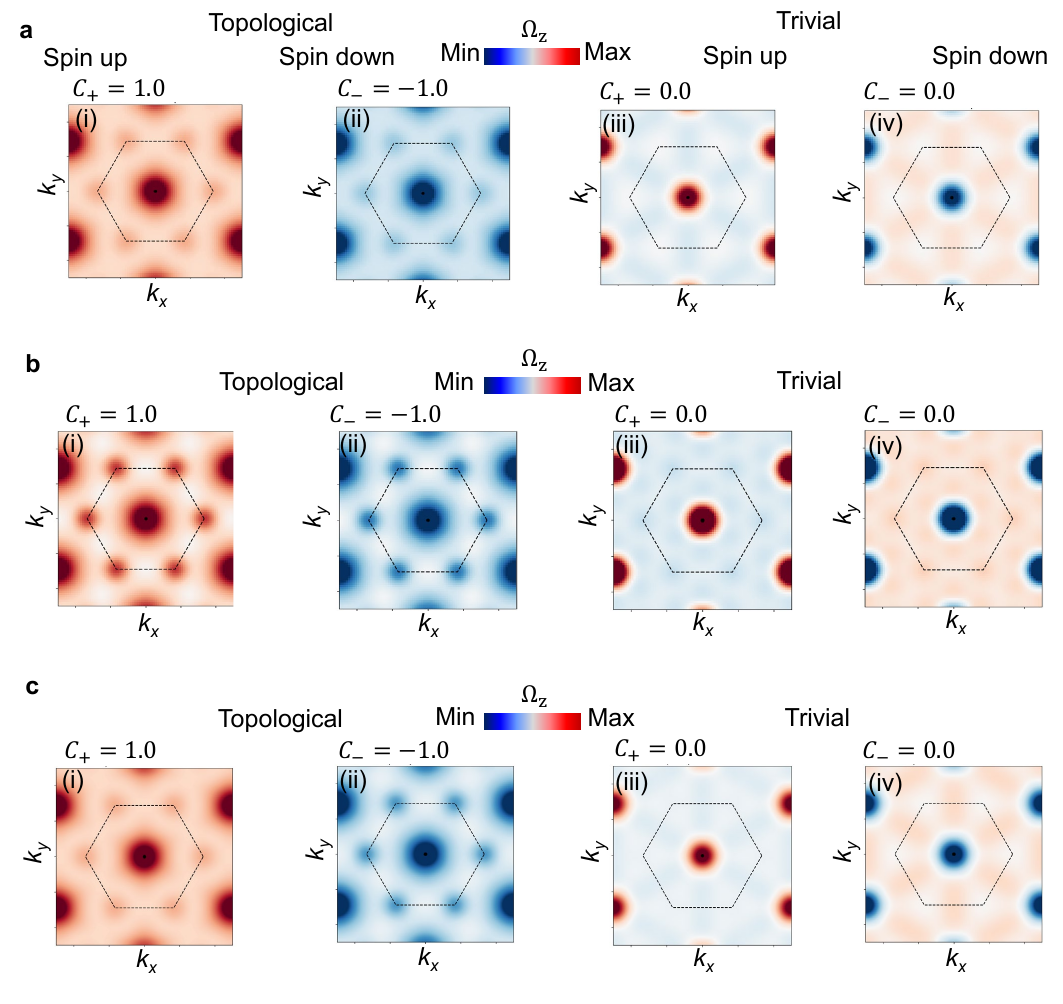}
	\caption{
    Berry phase maps in terms of the (\textbf{a}) first, (\textbf{b}) second, and (\textbf{c}) third bulk band gaps, respectively.
    The two-left and two-right panels show the Berry phase maps for topological and trivial phases.
    $C_{+(-)}$ represents the Chern number of the up- (down-)pseudospin sector.
    }
	\label{fig:figureS2}
\end{figure}

Spin Chern number, which characterizes the topological features in the quantum spin Hall (QSH)-based unit cell, is defined for each band gap by all occupied bands.
Here, we calculate spin Chern number by using 3, 9, and 15 bands below the first, second, third bulk band gaps, respectively, in Fig.~1d.
Figures~\ref{fig:figureS2}a-\ref{fig:figureS2}c show the 2D Berry phase maps for the first, second, and third complete gaps of the unit cell with $N=3$, respectively.
When the unit cell is designed as the topological phase (left two panels in Figs.~\ref{fig:figureS2}a-\ref{fig:figureS2}c), i.e. $k_{\mathrm{A}}<k_{\mathrm{B}}$, we find that pseudo spin-up and spin-down exhibit similar profiles for all three band gaps, resulting in the non-zero spin Chern number of $C_{\mathrm{S}}=(C_+-C_-)/2=1.0$.
On the other hand, the Berry phases for the trivial state, i.e. $k_{\mathrm{A}}>k_{\mathrm{B}}$, indicate $C_{\mathrm{S}}=C_+=C_-=0$ for each band gap.
Forming the topological boundary, the number of topological interface modes is determined by the difference in $C_{\mathrm{S}}$ for the topological and trivial phases, i.e. $\Delta C_{\mathrm{S}}$.
Thus, a helical pair of topological interface modes emerges in each bulk band gap by connecting the topological and trivial unit cells with $N=3$ because the topological and trivial unit cells exhibit $\Delta C_{\mathrm{S}}=1.0$ at all three bulk band gaps.

\clearpage

\section{Section 3: Effect of structural parameters of hierarchical components on dispersion diagrams}

As shown in our manuscript, the topological and trivial phases are determined by the ratio of intra- and inter-coupling, i.e. $k_{\mathrm{A}}$ and $k_{\mathrm{B}}$.
In this supporting section, we systematically show that the band closing for $k_{\mathrm{A}}=k_{\mathrm{B}}$ is maintained regardless of the parameters of other structural parameters.

\begin{figure}[h!]
	\centering
	\includegraphics[width=1.0\textwidth]{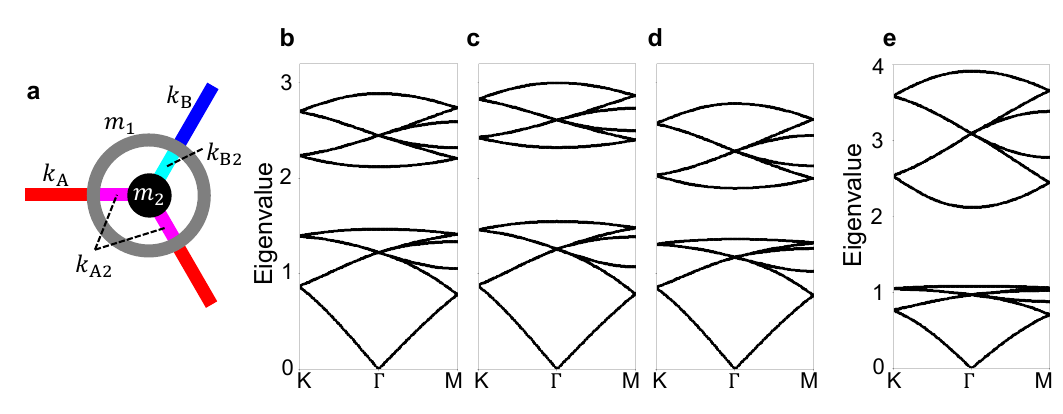}
	\caption{
    \textbf{a} Enlarged view of the sub-lattice in the unit cell with $N=2$.
     $k_\mathrm{A}$ is the intra-connections.
     $k_\mathrm{B}$ is the inter-connections.
     $k_\mathrm{A2}$ is the inner-connections parallel to $k_\mathrm{A}$.
     $k_\mathrm{B2}$ is the inner-connection parallel to $k_\mathrm{B}$.
     $m_1$ and $m_2$ are the masses of the host framework and inner resonator, respectively.
     \textbf{b-e} Dispersion diagrams for fixing $k_\mathrm{A}=k_\mathrm{B}=1.0$.
     In (\textbf{b}), other parameters are set to be $k_\mathrm{A2}=k_\mathrm{B2}=0.5$, $m_1=1.0$, and $m_2=0.5$.
     The inner connections for (\textbf(c)) are set to be $k_\mathrm{A2}=0.8$ and $k_\mathrm{B2}=0.2$.
     The inner connections for (\textbf(d)) are set to be $k_\mathrm{A2}=0.2$ and $k_\mathrm{B2}=0.8$.
     The dispersion diagram in (\textbf{e}) is obtained by setting $m_1=0.5$ and $m_2=1.0$ with fixing $k_\mathrm{A2}=k_\mathrm{B2}=0.5$.
        }
	\label{fig:figureS3}
\end{figure}

As the fundamental model, we verify the effect of the parameters on dispersion diagrams by using the honeycomb lattice with $N=2$.
Figure~\ref{fig:figureS3}a shows the enlarged view of one of the sub-lattices, where the intra-coupling (red lines), inner-coupling (blue line), two-inner coupling (magenta lines), another inner coupling (cyan line), host framework (gray ring), and innermost resonator (black circle) are expressed as $k_{\mathrm{A}}$, $k_{\mathrm{B}}$, $k_{\mathrm{A2}}$, $k_{\mathrm{B2}}$, $m_1$, and $m_2$, respectively.
For the dispersion diagrams in Figs.~\ref{fig:figureS3}b-\ref{fig:figureS3}d, we fix with $k_{\mathrm{A}}=k_{\mathrm{B}}=1.0$, $m_1=1.0$, and $m_2=0.5$.
First, Figure~\ref{fig:figureS3} shows the dispersion diagram for $k_{\mathrm{A2}}=k_{\mathrm{B2}}=0.5$, which is a similar condition to the unit cell with $N=2$ in the main manuscript.
Thus, we can see the two band closing at $\Gamma$ point.
Secondly, we show the dispersion diagrams for $k_{\mathrm{A2}}>k_{\mathrm{B2}}$ ($k_{\mathrm{A2}}=0.8$, $k_{\mathrm{A2}}=0.2$) and $k_{\mathrm{A2}}<k_{\mathrm{B2}}$ ($k_{\mathrm{A2}}=0.2$, $k_{\mathrm{A2}}=0.8$) in Figs.~\ref{fig:figureS3}c and \ref{fig:figureS3}d, respectively.
The band positions vary by the asymmetric inner couplings, while the two band closing points are maintained. 

Finally, we change the parameters to be $m_1<m_2$ ($m_1=0.5$ and $m_2=1.0$), and fix to be $k_{\mathrm{A2}}=k_{\mathrm{B2}}=0.5$.
The obtained dispersion diagram is shown in Fig.~\ref{fig:figureS3}e.
In the condition, there are also two band closing points regardless of the distortion of the band shapes.
Therefore, we find that the parameters of the hierarchical components do not affect the band opening, in other words, the topological features are independent of the inner hierarchical structures in the tight-binding analysis.

\clearpage

\section{Section 4: Numerical simulation of topological and trivial unit cells with $N=2$ for device fabrication}

\begin{figure}[h!]
	\centering
	\includegraphics[width=1.0\textwidth]{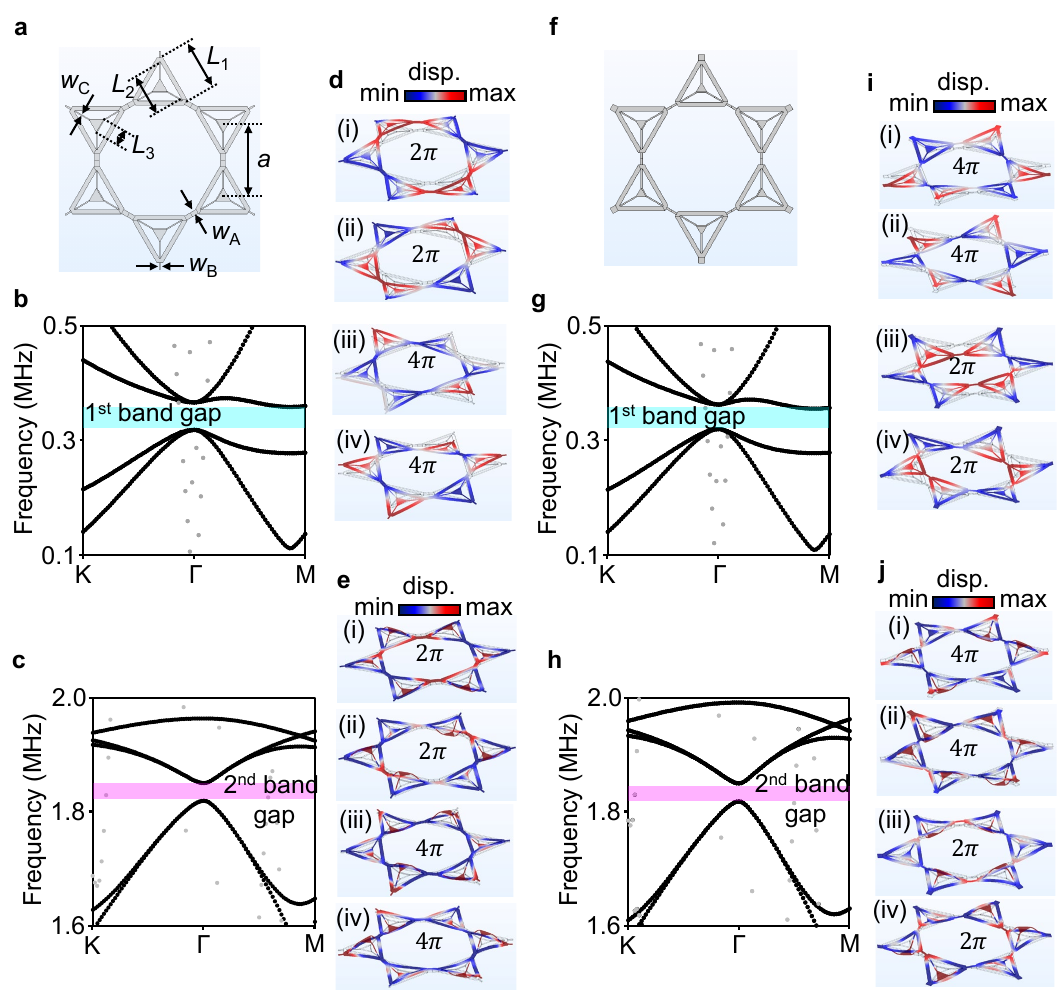}
	\caption{
    \textbf{a} Numerically designed model of the topological unit cell with $N=2$.
    \textbf{b, c} Numerical dispersion diagrams around the (\textbf{b}) first band gap (shaded cyan region) and (\textbf{c}) second band gap (shaded magenta region).
    The black and gray plots represent the transverse and longitudinal wave modes.
    \textbf{d, e} Displacement profiles in the topological unit cell at the modes (i and ii) above and (iii and iv) below the (\textbf{d}) first and (\textbf{e}) second band gaps.    
    \textbf{f} Numerically designed model of the trivial unit cell with $N=2$.
    \textbf{g, h} Numerical dispersion diagrams around (\textbf{g}) the first band gap (shaded cyan region) and (\textbf{h}) the second band gap (shaded magenta region).
    The black and gray plots represent the transverse and longitudinal wave modes.
    \textbf{i, j} Displacement profiles in the trivial unit cell at the modes (i and ii) above and (iii and iv) below the (\textbf{i}) first and (\textbf{j}) second band gaps.    
    }
	\label{fig:figureS4}
\end{figure}

We numerically design the unit cell to experimentally demonstrate the topological waveguide consisting of the honeycomb lattices for $N=2$.
Figure~\ref{fig:figureS4}a shows the top overview of the numerical model for the topological unit cell.
Each sub-lattice has the triangle host framework and inner resonator, which are designed with $L_1=50$ \textmu m, $L_2=38$ \textmu m, $L_3=13$ \textmu m, and $a=60$ \textmu m.
For the topological phases, the intra-beam, inner-beams, and innermost beam have widths of $w_{\mathrm{A}}=3.8$ \textmu m, $w_{\mathrm{B}}=1.2$ \textmu m, and $w_{\mathrm{C}}=0.8$ \textmu m, respectively.
The numerical model has a thickness of 1 \textmu m, which is consistent with the thickness of the top silicon layer.

Figures~\ref{fig:figureS4}b and \ref{fig:figureS4}c show the dispersion diagrams around the first and second band gaps (shaded cyan and magenta regions). 
Figures~\ref{fig:figureS4}d and \ref{fig:figureS4}e indicate the mode profiles for (i and ii) above and (iii and iv) below doubly degenerated modes at $\Gamma$ point in Figs.~\ref{fig:figureS4}b and \ref{fig:figureS4}c, respectively.
The modes classified from (i) to (iv) exhibit similar mode profiles in the host framework for both bulk band gaps.
On the other hand, the internal resonators operate in-phase and out-of-phase with the host framework at the first and second bands, respectively.

For the trivial unit cell as shown in Fig.~\ref{fig:figureS4}f, the structural parameters are set to be $L_1=51$ \textmu m, $L_2=39$ \textmu m, $L_3=14$ \textmu m, $w_{\mathrm{A}}=1.65$ \textmu m, $w_{\mathrm{B}}=5.3$ \textmu m, and $w_{\mathrm{C}}=1.0$ \textmu m.
The thickness and center-to-center distance between the nearest neighboring sub-lattices are 1 \textmu m and 60 \textmu m, respectively, similar to those for topological unit cell.

Figures~\ref{fig:figureS4}g and \ref{fig:figureS4}h show the dispersion diagrams around the first and second band gaps (shaded cyan and magenta regions) for the trivial unit cell. 
We see that the two bulk band gaps for trivial phase are designed in the similar frequency regions to those for topological phase. 
Thus, combining the topological and trivial unit cells, we obtain the topological boundary having two distinct topological interface modes as shown in the main manuscript. 
Figure \ref{fig:figureS4}i (\ref{fig:figureS4}j) shows the displacement profiles at the doubly degenerated modes (i and ii) above and (iii and iv) below the first (second) band gap at $\Gamma$ point in Fig.~\ref{fig:figureS4}g (\ref{fig:figureS4}j), respectively.
We find that the host structures at the mode from (i) to (iv) in the second band gap exhibit similar displacement profiles to those at the modes of the same classification in the first band gap.
Furthermore, the order of the displacement profiles in Figs.~\ref{fig:figureS4}d, \ref{fig:figureS4}e, \ref{fig:figureS4}i, and \ref{fig:figureS4}j evidently show that the bands inversions occur between the trivial and topological unit cells at both band gaps, indicating the topological phase transition.

\clearpage

\section{Section 5: Numerical simulation for the conventional second-order topological interface modes}

\begin{figure}[h!]
	\centering
	\includegraphics[width=1.0\textwidth]{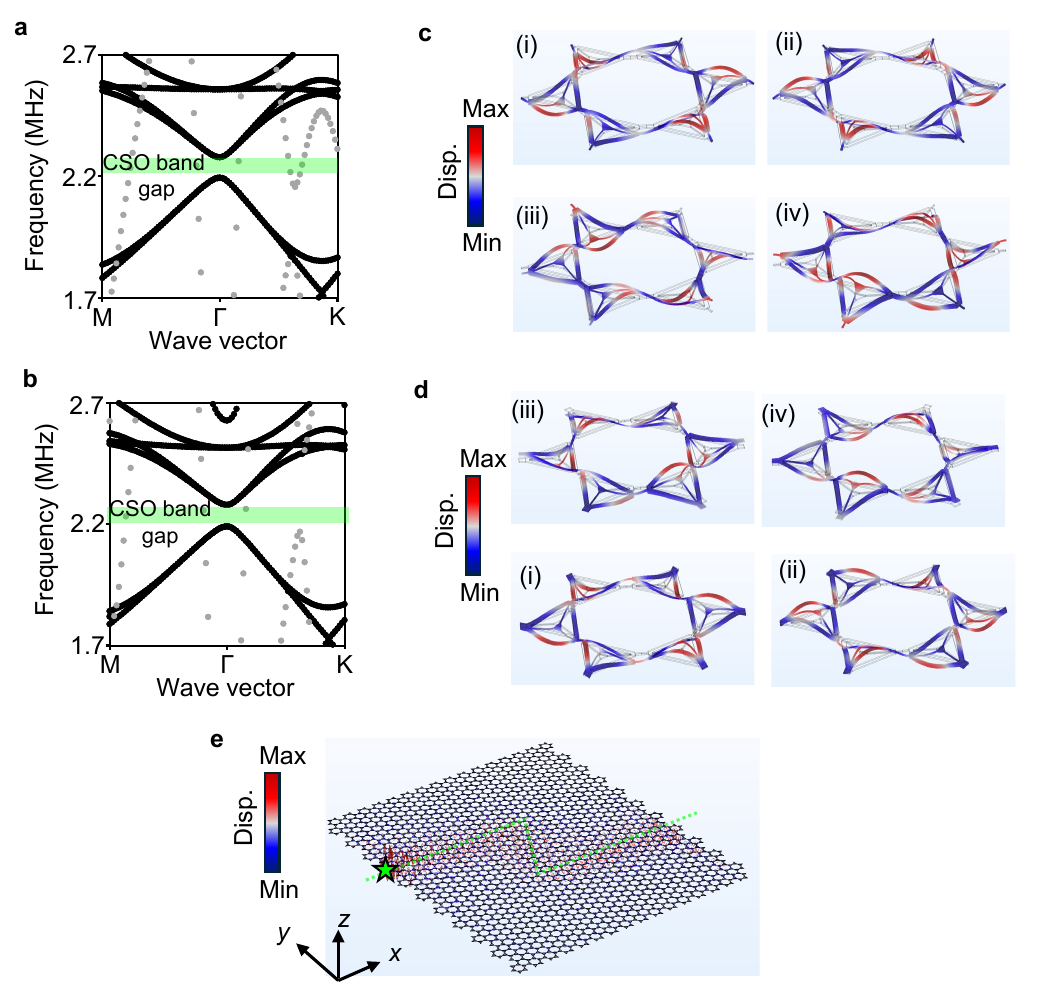}
	\caption{
    \textbf{a, b} Dispersion diagrams of (\textbf{a}) topological and (\textbf{b}) trivial unit cells around the conventional second-order (CSO) bulk band gap (shaded green region).
    The black and gray plots represent transverse and longitudinal wave modes.
    \textbf{c, d} Displacement profiles for (\text{c}) topological and (\textbf{d}) trivial unit cells at the two doubly degenerated modes above and below the CSO bulk band gap.
    The classified modes as (i, ii) and (iii, iv) are the higher-order p- and d-like modes with phase winding of $2\pi$ and $4\pi$.
    \textbf{e} Mode profile on the silicon membrane having the Z-shaped topological waveguide at an excitation frequency of 2.26 MHz corresponding to the CSO topological interface mode.
    The green star and line represent the excitation point and topological boundary.
    }
	\label{fig:figureS5}
\end{figure}

The numerical models in section 4 have a conventional second-order (CSO) bulk band gap, which is based on the complex spatial profiles.
Figures~\ref{fig:figureS5}a and \ref{fig:figureS5}b show the dispersion diagrams around the CSO bulk band gap (shaded green region) for topological and trivial unit cells, respectively.
Note that the gray plots in Figs.\ref{fig:figureS5}a and \ref{fig:figureS5}b represent the longitudinal wave modes, which are ignorable when the 2D structure is excited in the out-of-plane direction.
Figure~\ref{fig:figureS5}c [\ref{fig:figureS5}d] shows the displacement profiles at the two doubly degenerated modes (i, ii) above [below] and (iii, iv) below [above] the CSO bulk band gap.
We see that the modes classified as (i, ii) and (iii, iv) correspond to the higher-order p- and d-like modes with phase winding of $2\pi$ and $4\pi$, respectively.
Therefore, connecting the topological and trivial unit cells, we can observe the CSO topological interface modes in the single-line supercell as shown in Fig.~4a.

Furthermore, Figure~\ref{fig:figureS5}e shows the profile of the topological interface mode on the 2D structure at 2.26 MHz.
The structure is excited from the left edge of the waveguide.
We find that the excited elastic wave propagates along the Z-shaped topological boundary.
Therefore, we clearly show that the CSO topological interface modes also exhibit topological protection.

\clearpage

\section{Section 6: Numerical model of the topological unit cell with $N=3$}

\begin{figure}[h!]
	\centering
	\includegraphics[width=1.0\textwidth]{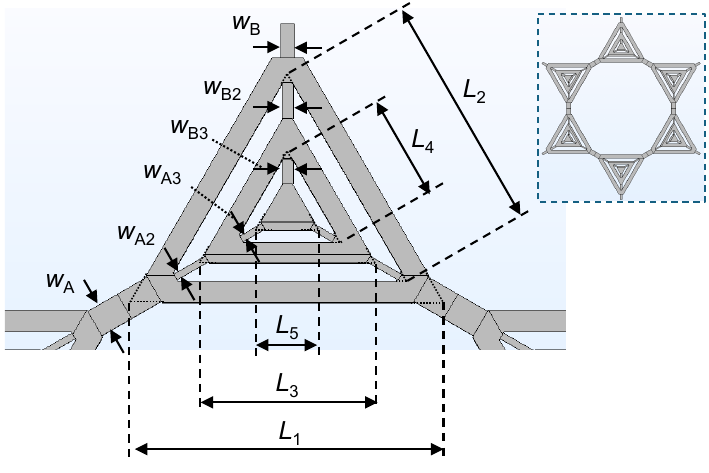}
	\caption{
    Numerical model of the sub-lattice in the topological unit cell with $N=3$.
    The inset represents the top overview of the unit cell.
    }
	\label{fig:figureS6}
\end{figure}

Figure~\ref{fig:figureS6} shows the enlarged view of one of the numerical sub-lattices model for the topological unit cell with $N=3$.
The inset shows the top view of the topological unit cell.
As described in the main manuscript, each of triangular masses and beams can have a different thickness.
In addition to the thicknesses, we tune other structural parameters to design the clear three bulk band gaps in Fig.~5c.
The triangular host structure, middle resonator, and innermost resonator are designed with the parameters of $L_1=50$ \textmu m and $L_2=38$ \textmu m, $L_3=28$ \textmu m and $L_4=17$, and $L_5=10$ \textmu m, respectively.
The intra- and inter-beam widths are $w_{\mathrm{A}}=4.5$ \textmu m and
$w_{\mathrm{B}}=2.2$ \textmu m, respectively.
In the model, we individually design the inner beams parallel to the intra- and inter-beams with $w_\mathrm{A2}=w_\mathrm{A3}=1.05$ \textmu m and $w_\mathrm{B2}=w_\mathrm{B3}=1.8$ \textmu m to fine-tune each band gap.

